\documentclass[letterpaper,12pt]{article}   
\usepackage{osajnl2} 

\begin{document}

\title{Energy transfer between two filaments due to degenerate four-photon parametric process}

\author{Daniela A. Georgieva $^{1}$ and Lubomir M. Kovachev $^{2,*}$}
\address{$^1$ Faculty of Applied Mathematics and Computer Science,
Technical University of Sofia, 8, Kliment Ohridski Blvd., 1000
Sofia, Bulgaria}
\address{$^2$ Institute of Electronics, Bulgarian Academy of Sciences,\\
Tzarigradcko shossee 72,1784 Sofia, Bulgaria}
\address{$^*$ Corresponding author: lubomirkovach@yahoo.com}

\begin{abstract}
Recently energy exchange between two filaments crossing at small
angle and with power slightly above the critical for self-focusing
$P_{cr}$ was experimentally demonstrated. In this paper we present a
model describing the process of this transfer through degenerate
four-photon parametric mixing. Our model confirms the experimental
results that the direction of energy exchange depends on the
relative transverse velocity (incident angle), laser intensity and
initial distance between the pulses (relative initial phase). We
also investigate the interaction between two collinear filaments in
order to explain the filaments number reduction for powers close to
$P_{cr}$ in multi-filamental propagation.
\end{abstract}

\ocis{190.4380, 320.2250}
\maketitle

\section{Introduction}
Several experimental works \cite{BERN}-\cite{DING} report intensive
exchange of energy when two filaments with the same carrying
frequency intersect at a small angle. The mechanisms for energy
exchange proposed in references \cite{BERN} and \cite{LIU} require
two different main frequencies of the pulses. The experiment
described in \cite{DING} is performed at a single pulse frequency.
The authors of \cite{DING} add another mechanism as a possible
explanation - nonlinear absorption at high input power. This
process, though, cannot lead to a periodical energy exchange. All
experiments were performed with filament intensities slightly above
the critical value for self-focusing $I_{cr}$. The plasma density
for laser pulses with such intensities is not high enough to provide
the effect of plasma grating. The  stability and the interaction of
filaments with intensities of the order of  $I\sim 10^{12}$ $W/cm^2$
some authors try to explain as balance between Kerr self-focusing
and ionization induced defocusing. In the book of P. Gibbon
\cite{GIBON} the ionization induced defocusing is discussed widely.
There are full coincidences between the numerical and experimental
results, cited in the book. It is shown, that the ionization induced
defocusing is possible only for intensities of the order of $I\sim
10^{15}$ $W/cm^2$, while in the experiments the  measured intensity
of a stable filament is of the order of $I\sim 10^{12}$ $W/cm^2$ -
which is three orders of magnitude less. This is the reason to look
for other mechanisms leading to exchange of energy and connected
with nonlinear optical effects. To do that, we take into account the
following experimental details. First, the laser pulses propagating
on different optical trajectories always intersect with a relative
phase difference. This means that at the intersection region there
are conditions for degenerate four-photon parametric (FPP)
processes. As is mentioned in \cite{BOYD}, the degenerate FFP
processes play an important role also in the multi-component
filamentation generation. Another important experimental observation
in the multi-filament propagation is the significant reduction in
the number of filaments as a function of the distance \cite{CHIN,
COUA, KILO}. Up to now we have not seen a physical model explaining
this phenomenon.

In this manuscript we propose a nonlinear vector model including
degenerate FFP process. We investigate numerically the interaction
between optical pulses in the cases when 1) the components of the
polarization vector are separated, 2) the filaments contain the two
polarization components (the polarization components are not
separated, general case). The phenomenon of the reduction of the
number of filaments during multi-filament propagation is also
studied in our model. Our analysis shows that the decreasing of the
number of filaments could be due to pairing of the filaments and
energy exchange due to degenerate FFP mixing. The numerical scheme
is performed on the base of the split-step Fourier method.

\section {Nonlinear Polarization}

As  demonstrated in \cite{BRAUN},  high intensity femtosecond pulses
generate in gases stable filaments with broad-band spectrum .
Various models explaining this phenomenon are presented (see for
example the paper of Couairon and Mysyrowicz and references therein
\cite{COUA}). The standard filamentation model  is based on plasma
generation and multi-photon processes and includes also nonlinear
polarization of the kind

\begin{eqnarray}
\label{NLKERR} \vec{P}^{nl} = n_2 \left[\left( \vec{E} \cdot
\vec{E}^* \right)\vec{E} + \frac{1}{2} \left( \vec{E} \cdot \vec{E}
\right)\vec{E}^*\right],
\end{eqnarray}
where $n_2$ is the nonlinear refractive index of the isotropic
media.  The polarization (\ref{NLKERR}) was proposed by Maker and
Terhune in 1965  \cite{MAKER}. If the electrical field contains one
linear or circular component, the polarization (\ref{NLKERR})
describes only the self-action effect, while in the case of
two-component electrical filed $\vec{E} = (E_x, E_y, 0)$ additional
terms appear, presenting cross-modulation and degenerate four-photon
parametric mixing. The self-action process broadens the pulse
spectrum - starting from narrow-band pulse, the stable filament
becomes broad-band far way from the source. Later \cite{KOL,LMK1}
show, though, that the  evolution of broad-band pulses like
filaments can not be described correctly by nonlinear polarization
of the kind (\ref{NLKERR}). It is more correct to use the
generalized nonlinear operator

\begin{eqnarray}
\label{NLTH} \vec{P}^{nl} = n_2 \left( \vec{E} \cdot \vec{E}
\right)\vec{E},
\end{eqnarray}
which includes additional processes associated with third harmonic
generation (THG). The more precise analysis, presented in the
present paper, demonstrates that the polarization of kind
(\ref{NLTH}) is not applicable to a scalar model, because the
corresponding Manley-Rowe (MR) conservation laws are not satisfied.
That is why we substitute into the nonlinear operators
(\ref{NLKERR}) and (\ref{NLTH}) a two-component electrical vector
field at one carrying frequency of the form:

\begin{eqnarray}
\label{ELF} \vec{E} =\frac{ \left( A_x \exp(i\omega_0 t) + c.c.
\right)}{2}\vec{x} + \frac{ \left( A_y \exp(i\omega_0 t) + c.c.
\right)}{2}\vec{y},
\end{eqnarray}
where $A_x=A_x(x, y, z, t), A_y = A_y(x, y, z, t)$ are the amplitude
functions and $\omega_0$ is the carrying frequency of the laser
source.

In the case of Maker and Terhune polarization (\ref{NLKERR}) we
obtain

\begin{eqnarray}
\label{POLKERR} \vec{P}^{nl}_x = \frac{3}{8} n_2 \left[ \left(
|A_x|^2 + \frac{2}{3}|A_y|^2\right) A_x + \frac{1}{3} A_x^* A_y^2
\right]\exp(i\omega_0 t) + c.c. \nonumber\\
\\
\vec{P}^{nl}_y = \frac{3}{8} n_2 \left[ \left( |A_y|^2 +
\frac{2}{3}|A_x|^2\right) A_y + \frac{1}{3} A_y^* A_x^2
\right]\exp(i\omega_0 t) + c.c. \nonumber
\end{eqnarray}
The nonrestricted nonlinear polarization (\ref{NLTH}) generates the
following components

\begin{eqnarray}
\label{POLTH} \vec{P}^{nl}_x = \frac{3}{8} n_2 \left[ \frac{1}{3}
\left( A_x^2 + A_y^2\right) A_x \exp(2 i \omega_0 t) + \left(
|A_x|^2 + \frac{2}{3}|A_y|^2\right) A_x + \frac{1}{3} A_x^* A_y^2
\right]\exp(i\omega_0 t) + c.c. \nonumber\\
\\
\vec{P}^{nl}_y = \frac{3}{8} n_2 \left[ \frac{1}{3} \left( A_x^2 +
A_y^2\right) A_y \exp(2 i \omega_0 t) + \left( |A_y|^2 +
\frac{2}{3}|A_x|^2\right) A_y + \frac{1}{3} A_y^* A_x^2
\right]\exp(i\omega_0 t) + c.c. \nonumber
\end{eqnarray}

Comparing (\ref{POLKERR}) and (\ref{POLTH}), it is clearly seen that
the operator (\ref{NLTH}) $ n_2 \left( \vec{E} \cdot \vec{E}
\right)\vec{E}$ generalizes the case of Marker and Terhune's
operator (\ref{NLKERR}), but includes also additional terms
associated with THG.

\section {Basic  System of Equations}

The stable filament propagation in gases is realized in the sub-pico
and femtosecond regions, while in the nano- and picosecond regions
appears the well known self-focusing. The dynamics of narrow-band
laser pulses can be accurately described in the frame of paraxial
optics. The filamentation experiments demonstrate a typical
 pulse spectrum evolution. The initial laser pulse
$\left( t_0\geq 50 fs \right) $ possesses a relatively narrow-band
spectrum $\left(\Delta k_z \ll k_0 \right)$, where $\Delta k_z$ is
the spectral pulse width and $k_0$ is the carrying wave number.
During the filamentation process the initial self-focusing broadens
significantly the pulse spectrum. The broad-band spectrum
$\left(\Delta k_z \sim k_0 \right)$ is one of the basic
characteristics of the stable filament. The evolution of the so
obtained filament can not be further described in the frame of the
nonlinear paraxial optics because the paraxial optics works
correctly for narrow-band laser pulses only. The dynamics of
broad-band pulses can be presented properly within different
non-paraxial models  such as UPPE  \cite{KOL} or non-paraxial
envelope equations \cite{KOVBOOK}. Another standard restriction in
the filamentation theory is the use of one-component scalar
approximation of the electrical field $\vec{E}$. This approximation
though,  is in contradiction with recent experimental results, where
rotation of the polarization vector is observed \cite{ZIG}. For this
reason, in the present paper we use the non-paraxial vector model up
to second order of dispersion, in which the nonlinear effects are
described by the nonlinear polarization components (\ref{POLTH}).
The system of non-paraxial equations of the amplitude functions
$A_x, A_y$ of the two-component electrical field (\ref{ELF}) has the
form

\begin{eqnarray}
\label{SYSGAL} -2i\frac{k_0}{v_{gr}} \frac{\partial A_x}{\partial
t}= \Delta_{\bot} A_x  - \frac{\beta + 1}{v_{gr}} \left(
\frac{\partial^2 A_x}{\partial t^2} - 2v_{gr} \frac{\partial^2
A_x}{\partial t
\partial z}\right) - \beta \frac{\partial^2 A_x}{\partial z^2}
\nonumber\\
+ k^2_0 \tilde{n}_2 \left[\frac{1}{3} \left(A^2_x + A^2_y
\right)A_x\exp\left(2ik_0\left(z-\left(v_{ph}-v_{gr}\right)t\right)\right)
+ \left( |A_x|^2 + \frac{2}{3}|A_y|^2\right) A_x + \frac{1}{3}
A^*_xA^2_y \right] \nonumber\\
\\
-2i\frac{k_0}{v_{gr}} \frac{\partial A_y}{\partial t}= \Delta_{\bot}
A_y - \frac{\beta + 1}{v_{gr}} \left( \frac{\partial^2 A_y}{\partial
t^2} - 2v_{gr} \frac{\partial^2 A_y}{\partial t
\partial z}\right) - \beta \frac{\partial^2 A_y}{\partial z^2}
\nonumber\\
+ k^2_0 \tilde{n}_2 \left[\frac{1}{3} \left(A^2_x + A^2_y
\right)A_y\exp\left(2ik_0\left(z-\left(v_{ph}-v_{gr}\right)t\right)\right)
+ \left( |A_y|^2 + \frac{2}{3}|A_x|^2\right) A_y+ \frac{1}{3}
A^*_yA^2_x \right], \nonumber
\end{eqnarray}
where $\tilde{n}_2 = \frac{3}{8} n_2$, $v_{gr}$ and $v_{ph}$ are the
group and phase velocities correspondingly, $\beta = k_0v_{gr}^2k''$
and $k''$ is the group velocity dispersion.

This model describes the ionization-free filamentation regime, where
the pulse intensities are close to the critical one for
self-focusing. The first nonlinear term in (\ref{SYSGAL})
corresponds to  coherent GHz generation \cite{LMK1}. The system
(\ref{SYSGAL}) is written in Galilean frame $\left( z'=z-vt; t'=t
\right)$ and not in the standard local time frame $\left(t'=t-z/v;
z'=z\right)$.  In all  coordinate systems - laboratory, moving in
time, and Galilean, the group velocity adds an additional phase
(carrier-envelope phase) in the third harmonic terms and transforms
them to GHz ones. This can be seen directly for the system
(\ref{SYSGAL}) written in Galilean frame, which determines the
choice of coordinates. The last nonlinear term in (\ref{SYSGAL})
describes degenerate four-photon parametric mixing (FPM).

The system of equations (\ref{SYSGAL}) written in dimensionless form
becomes
\begin{eqnarray}
\label{SYSGAL1} -2i\alpha\delta^2 \frac{\partial A_x}{\partial t}=
\Delta_{\bot} A_x  - \delta^2\left(\beta + 1\right) \left(
\frac{\partial^2 A_x}{\partial t^2} - \frac{\partial^2 A_x}{\partial
t \partial z}\right) - \delta^2\beta \frac{\partial^2 A_x}{\partial
z^2}
\nonumber\\
+ \gamma \left[\frac{1}{3} \left(A^2_x + A^2_y
\right)A_x\exp\left(2i\left(\alpha
z-\tilde{\omega}_{nl}t\right)\right) + \left( |A_x|^2 +
\frac{2}{3}|A_y|^2\right) A_x + \frac{1}{3}
A^*_xA^2_y \right] \nonumber\\
\\
-2i\alpha\delta^2 \frac{\partial A_y}{\partial t}= \Delta_{\bot} A_y
- \delta^2\left(\beta + 1\right) \left( \frac{\partial^2
A_y}{\partial t^2} - \frac{\partial^2 A_y}{\partial t
\partial z}\right) - \delta^2\beta \frac{\partial^2 A_y}{\partial z^2}
\nonumber\\
+ \gamma \left[\frac{1}{3} \left(A^2_x + A^2_y
\right)A_y\exp\left(2i\left(\alpha
z-\tilde{\omega}_{nl}t\right)\right) + \left( |A_y|^2 +
\frac{2}{3}|A_x|^2\right) A_y + \frac{1}{3} A^*_yA^2_x \right],
\nonumber
\end{eqnarray}
where $x = x/r_0$, $y = y/r_0$, $z = z/r_0$ are the dimensionless
coordinates, $r_0$ is the pulse waist, $z_0 = v_{gr}t_0$ is the
spatial pulse length, $\alpha=k_0z_0$, $\delta = r_0/z_0$, $\gamma =
k^2_0r^2_0\tilde{n}_2|A_0|^2/2$ is the nonlinear coefficient and
$\tilde{\omega}_{nl} = k_0\left(z-\left(v_{ph}-v_{gr}\right)\right)
t_0 $ is the normalized nonlinear frequency.

\section{Conservation Laws}
The nonlinear theories based on the polarization of Maker and Terhun
type (\ref{NLKERR}) satisfy the Manley-Rowe  relations. This means
that during the process of energy exchange the total energy is
conserved for arbitrary localized smooth complex fields. Additional
conservation quantities are also possible. To satisfy the MR
relations of the  truncated equations with a generalized nonlinear
polarization of type $\vec{P}^{nl} = n_2 \left( \vec{E} \cdot
\vec{E} \right)\vec{E}$,  some restrictions on the components of the
electrical field are imposed. We will demonstrate this on the basis
of two-component vector field
$\vec{E}=(A_x,A_y,0)\exp\left[ik_0(z-v_{ph}t)\right]$. Let us
rewrite the generalized nonlinear polarization (\ref{NLTH}) using
circularly polarized components \cite{BOYD, AGR}

\begin{eqnarray}
\label{CIRK} A_+ = (A_x+iA_y)/\sqrt{2}, \; A_- =
(A_x-iA_y)/\sqrt{2}.
\end{eqnarray}
Thus we obtain

\begin{eqnarray}
\label{NLCIRK} P_+ =
n_2\left(A_+^2A_-\right)\exp\left[ik_0(z-v_{ph}t)\right]\\
P_- = n_2 \left(A_-^2A_+\right)\exp\left[ik_0(z-v_{ph}t)\right],
\end{eqnarray}
where by convention $P_+$ and $P_-$ correspond to left-hand circular
and to right-hand circular polarization. The truncated equations can
be written as

\begin{eqnarray}
\label{TRUNC} i\frac{\partial A_+}{\partial t} = n_2\left(A_+^2A_-\right)\\
i\frac{\partial A_-}{\partial t}  = n_2 \left(A_-^2A_+\right).
\end{eqnarray}
The equations for the square modulus of the  components $A_+$ and
$A_-$ are

\begin{eqnarray}
\label{MODTRUNC} i\frac{\partial |A_+|^2}{\partial t} =
n_2|A_+|^2\left(A_+A_- - A^*_+A^*_-\right)=0\nonumber\\
\\
i\frac{\partial |A_-|^2}{\partial t} = n_2|A_-|^2\left(A_+A_- -
A^*_+A^*_-\right)=0,\nonumber
\end{eqnarray}
following from the fact that the  components $A_+$ and $A_-$ are
complex-conjugated fields (\ref{CIRK}). Thus we prove that to
satisfy the MR conditions for  the nonlinear system (\ref{SYSGAL1})
(or other conservative nonlinear equations) with nonlinear operator
of kind (\ref{NLTH}), the possible initial conditions and solutions
should be complex-conjugated fields. The conservation laws
(\ref{MODTRUNC}) give us additional information on the behavior of
the vector amplitude function: \emph{ only components of the vector
amplitude field $\vec{A}=(A_x,A_y,0)$, which present rotation of the
vector $\vec{A}$ in the plane $(x,y)$, satisfy the MR conditions}.
That is why in our numerical experiments, as well as in our
analytical investigations, we will use complex-conjugated components
only.

\begin{figure}
\centerline{\includegraphics[width=140mm,height=60mm]{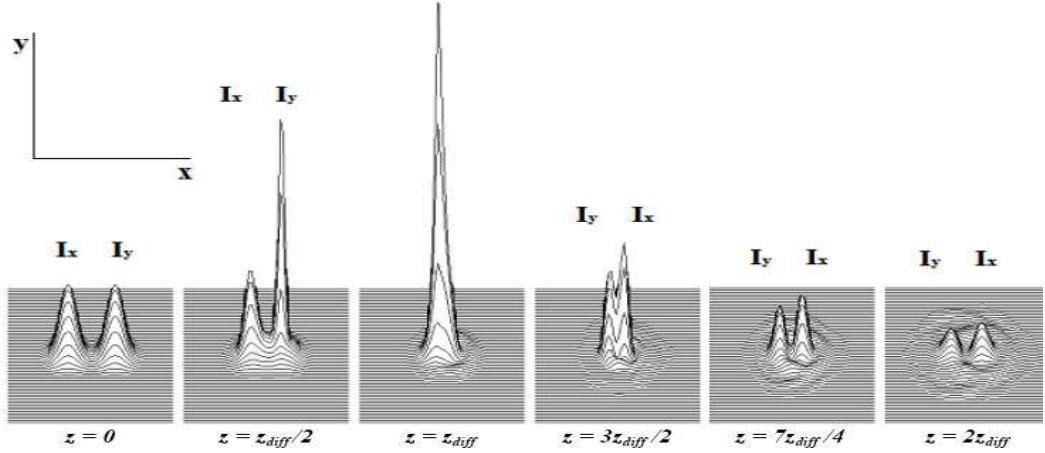}}
\caption{ Collision dynamics between two  separated polarization
components  at initial distance  $2a=4$, transverse velocity
difference $ 2\Delta v =3$, and nonlinear coefficient $\gamma=1,5$,
governed by the system of equations (\ref{SYSGAL1}) with initial
conditions (\ref{INCOND}). In the process of collision we observe
self-focusing and periodic exchange of energy. The initial  energy
transfer is from $I_x$ to $I_y$. With $z_{diff}$ is denoted the
diffraction lengths $z_{diff}=k_0r_0^2$.}
\end{figure}

\section{Numerical simulations and discussions}

We investigate numerically the following two basic nonlinear effects
in the  filamentation process - the energy exchange between two
non-collinear filaments and the phenomenon of reducing the number of
filaments in multi-filamentation propagation. We demonstrate that
both processes are based on degenerate FPM mixing. The numerical
simulations are carried out by using the split-step Fourier method.
The numerical results are presented for initial conditions:  $240
fs$ Gaussian bullet with waist and spatial length $r_0 = z_0 = 72
\mu m $ and power slightly above $P_{cr}$. In this case $\alpha =
200\pi$, $\delta = 1$,  $\tilde{\omega}_{nl}=0.00023$ and $\gamma
\in 1.5-3$. The phase difference between the two components is
initially $\pi/2$ in order to satisfy the conservation laws.

\begin{figure}
\centerline{\includegraphics[width=120mm,height=100mm]{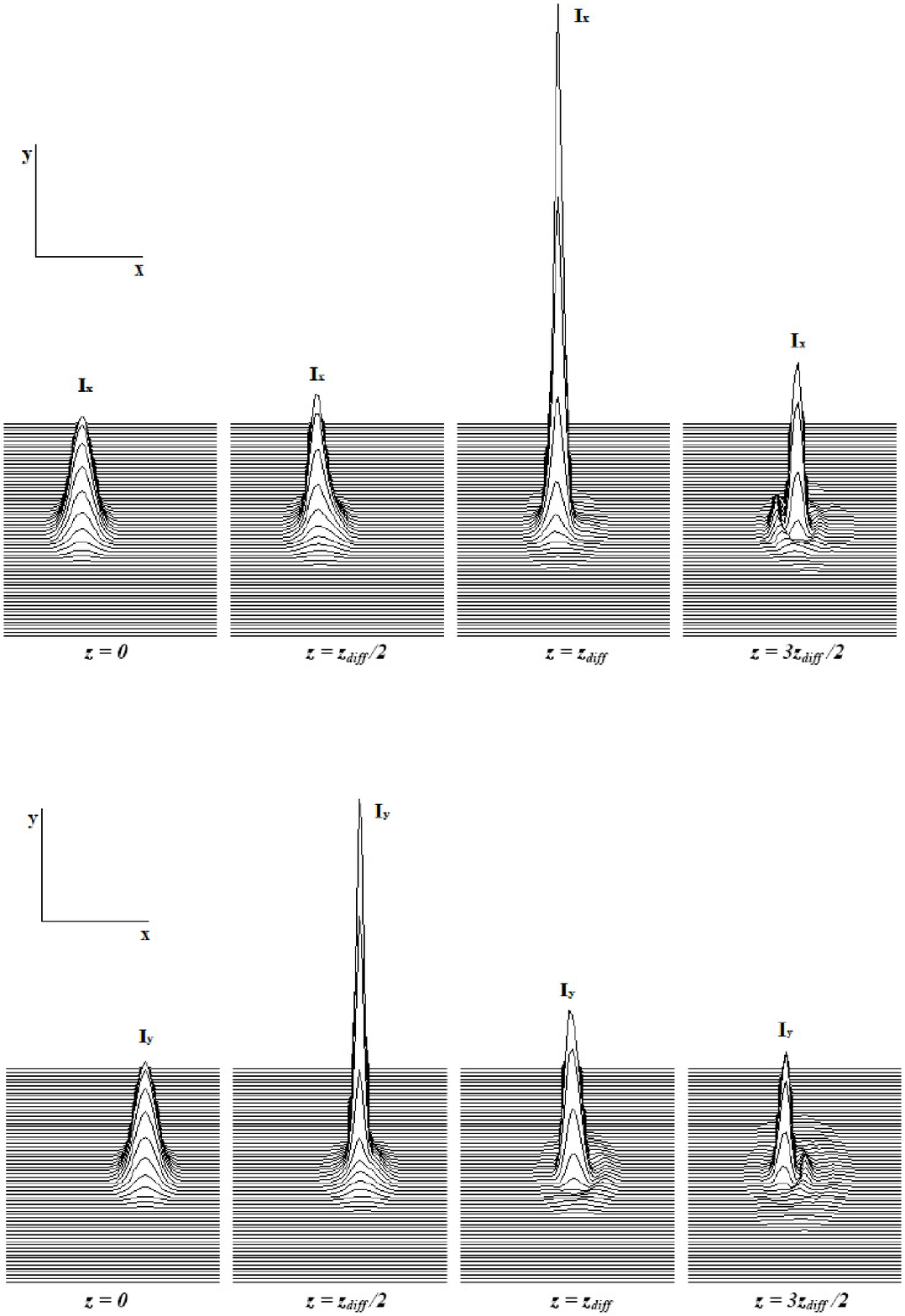}}
\caption{ Evolution of the intensity profile $I_x$ of the
$A_x$-component and the intensity profile $I_y$ of the
$A_y$-component of the vector field $\vec{E}$ (presented separately)
for the same initial conditions as in Fig. 1.  The periodicity of
the energy exchange is clearly seen.}
\end{figure}

\subsection{Energy Exchange between Two Non-collinear Filaments}
Let us begin our consideration with the case when two laser pulses
with the same carrying frequency intersect at a small angle. The
pulse power is slightly above the critical for self-focusing
$P_{cr}$.

\subsubsection{Polarization Separated Components}

The separation of the polarization components can be  realized
experimentally by means of a system of prisms, where the light pulse
falls at Brewster angle. In this case the initial conditions for
numerical solution of the system of equations (\ref{SYSGAL1}) have
the form

\begin{eqnarray}
\label{INCOND} A_x = A^0_x\exp\left(-\frac{(x+a)^2 + y^2 + z^2}{2}\right)\exp\left(-i\Delta vx\right)  \nonumber\\
\\
A_y=A^0_y\exp\left(-\frac{(x-a)^2 + y^2
+z^2}{2}\right)\exp\left(i\Delta vx\right)
\exp\left(i\frac{\pi}{2}\right), \nonumber
\end{eqnarray}
where $a$ is the initial shift of the pulses in $x$-direction with
respect to the intersection point and $2\Delta v = 2 v\sin\theta$ is
the normalized relative transverse velocity of the pulses ($\theta$
is the angle between the two trajectories).

The crossing  of the optical pulses $ A_x$ and $A_y$ for $\gamma =
1.5$, $\Delta v = 1.5$ and $a = 2$ is presented in Fig. 1. With
$I_x=|A_x|^2$ and $I_y=|A_y|^2$ are denoted the intensity profiles
of the spots ($x,y$) projections of the pulses, while $z_{diff}$
denotes the diffraction lengths $z_{diff}=k_0r_0^2$. The behavior of
the intensity profiles $I_x$ and $I_y$ separately is shown in Fig.
2. The intensive periodical energy exchange and simultaneous
self-focusing of $A_x$ and $A_y$ is evident. The periodicity of the
FFP process can be seen very well in Fig. 2. In this case the
initial energy transfer is from $I_x$ to $I_y$.

In our study the magnitude and direction of energy transfer depend
on the initial shift of the pulses $a$, the relative transverse
velocity $\Delta v$ and the laser intensity $\gamma$.  Fig. 3
presents a similar numerical simulation as  Fig. 1, with $\Delta v =
1$ (instead $\Delta v = 1.5$). In this experiment the direction of
energy transfer is changed from $I_y$ to $I_x$. This confirms the
experimental results obtained in \cite{DING}, where similar
dependencies of the transfer energy direction on the relative time
delay, laser intensities and intersecting angle ($\Delta v =
v\sin\theta$) are presented.

\begin{figure}
\centerline{\includegraphics[width=120mm,height=60mm]{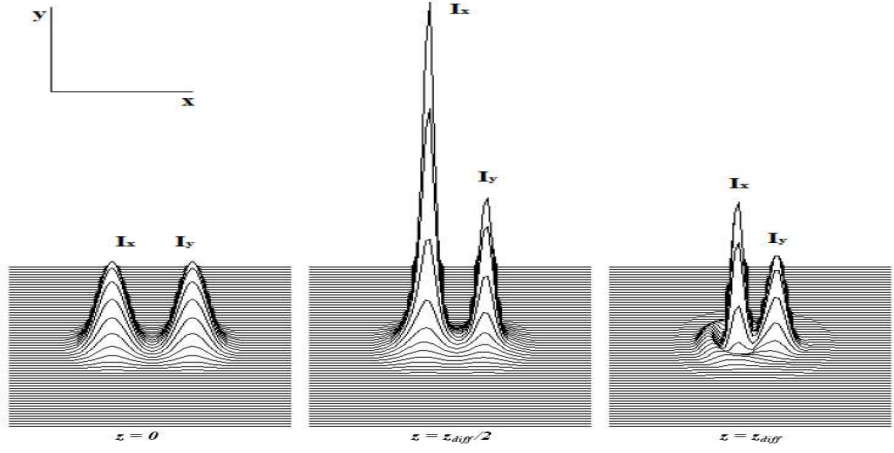}}
\caption{ Crossing dynamics between two  separated polarization
components  at initial distance  $2a=4$, transverse velocity
difference $2\Delta v =2$, and nonlinear coefficient $\gamma=1,5$,
governed by the system of equations (\ref{SYSGAL1}) with initial
conditions (\ref{INCOND}). The value of the parameter $2\Delta v$
changes the initial phase difference. As a result, the direction of
energy transfer is changed from $I_y$ to $I_x$.}
\end{figure}

\subsubsection{Non-separated Polarization Components}

When the laser pulse is split into two arms $\vec{A}_1$ and
$\vec{A}_2$ by  a regular beam splitter, the obtained filaments
contain each of the polarization components. Let $\vec{A}_j =
A_{j,x} \vec{x} + A_{j,y} \vec{y}, j = 1,2$. Then the initial
conditions for numerical solution of the system of equations
(\ref{SYSGAL1}) have the form

\begin{eqnarray}
\label{INCOND1}  A_x = A_{1,x} + A_{2,x} = \frac
{A^0_1}{\sqrt{2}}\exp\left(-\frac{(x+a)^2 + y^2 +
z^2}{2}\right)\exp\left(-i\Delta vx\right)  \nonumber \\
+ \frac{A^0_2}{\sqrt{2}}\exp\left(-\frac{(x-a)^2 + y^2
+z^2}{2}\right)\exp\left(i\Delta vx\right)
\exp\left(i\frac{\pi}{2}\right) \nonumber\\
\\
A_y = A_{1,y} + A_{2,y} =  \Biggr \{ \frac
{A^0_1}{\sqrt{2}}\exp\left(-\frac{(x+a)^2 + y^2 +
z^2}{2}\right)\exp\left(-i\Delta vx\right) \nonumber \\
+\frac{A^0_2}{\sqrt{2}}\exp\left(-\frac{(x-a)^2 + y^2
+z^2}{2}\right)\exp\left(i\Delta vx\right)
\exp\left(i\frac{\pi}{2}\right) \Biggr \}
\exp\left(i\frac{\pi}{2}\right), \nonumber
\end{eqnarray}
where $A_x$ and $A_y$ are composed of the $x$- and $y$-components of
the two optical pulses propagating along different trajectories.

The interaction of the optical pulses $\vec{A}_1$ and $\vec{A}_2$
for $\gamma = 1.5$, $\Delta v = 1.5$ and $a = 2$ is shown in Fig. 4.
As in the previous case of separated polarization components we
observe the effects of self-focusing and intensive periodical energy
exchange . In the faraway zone two additional components in
direction orthogonal to the initial pulses are obtained. We suppose
that these additional components are connected with the four-wave
mixing process. Recently similar experimental results have been
reported in \cite{WANG}.

\begin{figure}
\centerline{\includegraphics[width=150mm,height=60mm]{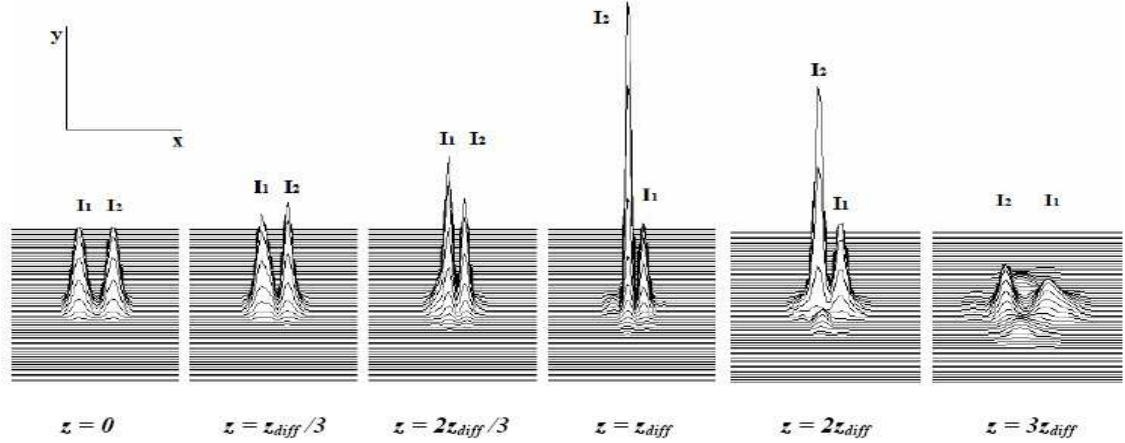}}
\caption{ Evolution of two filaments with non-separated polarization
components at initial distance  $2a=4$, transverse velocity
difference $ 2\Delta v =3$, and nonlinear coefficient $\gamma=1,5$,
governed by the system of equations (\ref{SYSGAL1}) with initial
conditions (\ref{INCOND1}). We observe self-focusing, periodic
exchange of energy and in the faraway zone - generation of two
additional components in direction orthogonal to the initial
pulses.}
\end{figure}

\subsection{Energy Exchange between Two Collinear Filaments}

The propagation through atmosphere of high intensity laser pulse
with power two orders of magnitude greater than the critical for
self-focusing ($P\sim 100 P_{cr}$) leads to breakup of the pulse
into many components, each with power  around $P_{cr}$ \cite{BOYD}.
The basic idea is that filamentation occurs as a consequence of
initially present laser wavefront irregularities, enhanced by
four-wave mixing. As  was reported recently \cite{KILO}, the number
of filaments is reduced significantly as a function of the distance.
We  propose a model based on the degenerate FFP process in order to
explain this decreasing number of filaments. Let us suppose that two
filaments are at a small distance from each other. Therefore there
are conditions for pairing and interaction due to degenerate FFP
mixing. The formulation of the problem is similar to the last case
in the previous subsection. Let us consider two filaments
$\vec{A}_1$ and $\vec{A}_2$ with arbitrary polarization. Let us
write the vectors through their $x$- and $y$-components $\vec{A}_j =
A_{j,x} \vec{x} + A_{j,y} \vec{y}, j = 1,2$. Then the initial
conditions for the numerical solution of the system (\ref{SYSGAL1})
have the form (\ref{INCOND}), with relative velocity $\Delta v =
0.00001$ (hence $\theta \sim 0$, i.e. the filaments are collinear).
\begin{figure}
\centerline{\includegraphics[width=130mm,height=60mm]{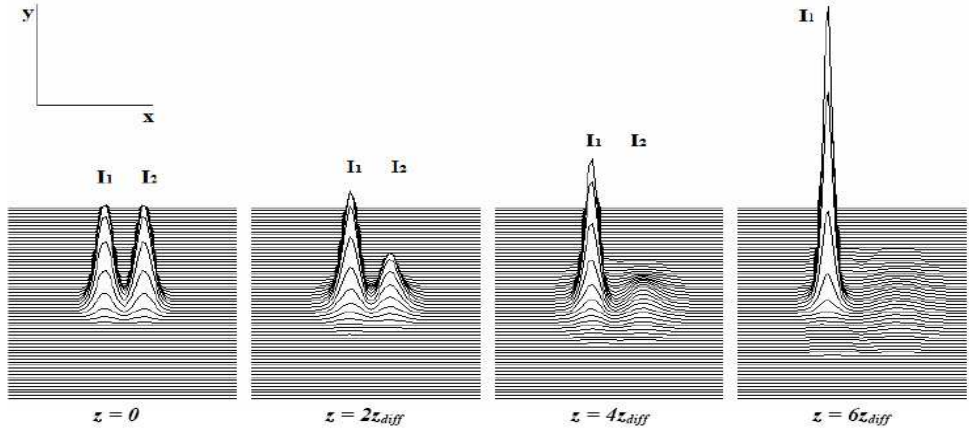}}
\caption{Energy exchange between two collinear filaments $\vec{A}_1$
and $\vec{A}_2$ at small a distance $2a=3.4$ governed by the system
of equations (\ref{SYSGAL1}) with initial conditions (\ref{INCOND1})
at values of the parameters $v=0.00001$ and $\gamma=1.5$. Due to
degenerated FFP mixing one of the filaments is amplified while the
other filament enters in linear mode and vanishes.}
\end{figure}

Fig 5. shows the evolution of two filaments and their exchange of
energy by degenerate FFP mixing. We use  power close to $P_{cr}$.
Therefore the amplified pulse self-focuses and gets enough power to
continue its propagation, while the other pulse gives out energy,
enters into  linear mode and vanishes. In this way the number of
filaments can be reduced by non-linear parametric processes in
$\chi^{(3)}$ media.

\section{Conclusions}

We have developed a model to describe the recent experimental
demonstration of energy exchange between two non-collinear filaments
in air, with power close to $P_{cr}$ \cite{BERN}-\cite{DING}. The
authors there point out a possible mechanism based on the generation
of plasma grating at the interaction point. It is known though, that
at higher incident laser power, where the plasma plays an  important
role, the exchange efficiency decreases. In air $P=P_{cr}$
corresponds to intensity of the laser field of the order of $I\sim
10^{12}$ $W/cm^2$ , where the plasma density is too small to form
plasma grating.

In this paper we propose another mechanism for this phenomenon on
the basis of degenerate FFP mixing. Our numerical investigation
confirms the experimental results of periodical energy exchange in
the region of overlapping of the two pulses. We obtain also the
observed in the experiments dependence of the initial energy
transfer on the relative transverse velocity (crossing angle),
intensity and initial phase difference (distance $a$). When the
pulse $A_1$ has positive initial phase  with respect  to pulse $A_2$
(depending on the above parameters), initially pulse $A_1$ takes
energy from pulse $A_2$ and vice versa. As  pointed out in
\cite{BOYD}, the degenerate FFP  processes play an important role in
the multi-filament generation. Another important effect, which was
observed experimentally in the multi-filament propagation, is that
the number of filaments $N$ with power arround $P_{cr}$ decreases
gradually with propagation distance \cite{CHIN, COUA, KILO}. The
authors explain this with energy losses by plasma or multi-photon
absorbtion. We think, though, that for intensities of the  order of
$I\sim 10^{12}$ $W/cm^2$ the multi-photon processes and plasma are
very weak to play such important role in this process. That is why
we turn back to the theory developed by Boyd \cite{BOYD}, and we
extend it now to evolution equations  for pairing of pulses by
degenerate FFP process. The result is that one of the pulses is
amplified, while the other one decreases in energy, enters in linear
mode and vanishes. Thus, the degenerate FFP process is a natural
explanation of  the reducing of the number of filaments with power
around $P_{cr}$. Another important result in this study is that we
investigate the filament and the filament's interaction as a vector
field. Beyond the scope of the specific task, we have shown that the
vector representation of the problems removes the disadvantages of
the use of the generalized nonlinear polarization operator
(\ref{NLTH}), making it a useful tool in the solution of different
problems, connected with third-order polarization. We plan to use in
the future this (or a similar) vector system for description of the
observed in  \cite{WANG} FFP vector solitons.

\section{ACKNOWLEDGMENTS}
This work was supported in part by Technical University of Sofia
under grant No.$141PD0001-11$ and grant DDVU02/71 with the Bulgarian
Science Fund.

\end{document}